\documentclass[12pt]{article}

\usepackage{graphicx}

\usepackage[tbtags]{amsmath}
\usepackage{chicago}
\usepackage{amsfonts}
\usepackage{amssymb}

\renewcommand{\eqref}[1]{(\ref{#1})}

\renewcommand{\vec}[1]{\mbox{\boldmath $#1$}}
\newcommand{\mat}[1]{{\mbox{\bfseries \rmfamily#1}}}
\newcommand{\Idmat}{\vec{1}}

\newcommand{\fave}{{\langle w \rangle}}
\newcommand{\piave}{{\bar \pi}}

\begin{document}

\title{Probability of fixation of an advantageous mutant in a viral quasispecies}
\author{Claus O. Wilke\\Digital Life Laboratory 136-93\\Caltech, Pasadena CA 91125}

\maketitle

\newpage

\noindent
Running head: Quasispecies fixation
\bigskip

\noindent
Keywords: quasispecies, fixation probability, RNA virus evolution, neutral
network
\bigskip

\noindent
Corresponding author:\\
\mbox{}~~Claus O. Wilke\\
\mbox{}~~Digital Life Laboratory 136-93\\
\mbox{}~~Caltech, Pasadena CA 91125\\
\mbox{}~~Tel.:\ 626 395 2338\\
\mbox{}~~Fax: 626 395 2944\\
\mbox{}~~wilke@caltech.edu

\bigskip

\noindent
{\bf Abstract:} The probability that an advantageous mutant rises to fixation
in a viral quasispecies is investigated in the framework of multi-type
branching processes. Whether fixation is possible depends on the overall
growth rate of the quasispecies that will form if invasion is successful,
rather than on the individual fitness of the invading mutant. The exact
fixation probability can only be calculated if the fitnesses of \emph{all}
potential members of the invading quasispecies are known. Quasispecies
fixation has two important characteristics: First, a sequence with negative
selection coefficient has a positive fixation probability as long as it has
the potential to grow into a quasispecies with an overall growth rate that
exceeds the one of the established quasispecies. Second, the fixation
probabilities of sequences with identical fitnesses can nevertheless vary  over
many orders of magnitudes. Two approximations for the probability of fixation
are introduced. Both approximations require only partial  knowledge about the
potential members of the invading quasispecies. The performance of these two
approximations is compared to the exact fixation probability on a 
network of RNA sequences with identical secondary structure.  \newpage

\centerline{INTRODUCTION}
\bigskip

\noindent
One of the most remarkable aspects of the dynamics of RNA viruses is the high
rate at which mutant variants are produced. At mutation rates close to one
substitution per genome per generation
\cite{Drake93,DrakeHolland99},
a virus population forms a highly diverse cloud of mutants
\shortcite{Domingoetal76,Domingoetal78,Hollandetal82,Steinhaueretal89,BiebricherLuce93,BurchChao2000},
a so-called quasispecies
\shortcite{EigenSchuster79,Nowak92,DomingoHolland97,Domingoetal2001}.
At the same time, the sequence space is so large that even for population
sizes up to $10^{12}$, there is a constant stream of new mutants that have
never existed before. Most of these mutants have impaired fitness, but
occasionally, a new mutant will fare better than all currently existing
virions, for example by presenting an epitope that the immune system fails to
recognize. With a certain probability, this mutant will rise to fixation,
where fixation is understood in the sense that the mutant becomes the ancestor
of a new quasispecies which completely replaces the currently existing one.

The problem of the fixation of an advantageous mutant is an old one, with a
long history of investigations in classical population genetics, reaching back
to Haldane and Fisher
\shortcite{Fisher22,Fisher30,Haldane27,Kimura57,Kimura64,Kimura70,KimuraKing79,Ewens67,BurgerEwens95,Barton95,OttoBarton97,Pollak2000}.
However, these investigations differ from the quasispecies case in one
important aspect: the mutation rates considered. In classical population
genetics, the usual assumption is that mutations are rare events, such that an
invading mutant will not mutate again while it is moving towards either
fixation or extinction. In the quasispecies setting, on the other hand, most
of the immediate offspring of a mutant will have further mutations, and their
offspring as well, and so on. As a consequence, the fitness of a prospective
invading quasispecies is not given by the fitness of the initial mutant, but
rather by the average fitness of the offspring mutant cloud that will form
eventually. One of the more surprising results of this dynamics is that a
mutant with the ability to replace the currently existing quasispecies may
actually have a reduced replication rate, if at the same time its robustness
against further mutations is increased
\shortcite{SchusterSwetina88,Wilkeetal2001b,Wilke2001b,KrakauerPlotkin2002}.

Quasispecies theory in its original formulation by Eigen and Schuster
\citeyear{EigenSchuster79} is based on deterministic differential equations,
and as such cannot deal with the fluctuations that are responsible for
fixation or extinction of individual mutants.  Within the more general
mathematical framework of multi-type branching processes, it is possible to
describe both the deterministic aspects of large populations as well as the
fluctuations inherent in the dynamics of small and very small populations
\shortcite{Demetriusetal85,HofbauerSigmund88,Hermissonetal2002}.  An
expression for the probability of fixation follows naturally from branching
process theory. We will discuss how this expression relates to the predictions
of the deterministic quasispecies equations, as well as to the results of
classical population genetics.

The remainder of the paper is organized as follows. First, we derive a general
expression for the probability of fixation in an arbitrary fitness landscape.
Then, we discuss the special case of fixation on a neutral network, that is,
the case in which all sequences of the invading quasispecies have the same
fitness, and derive two approximations for the fixation probability that can
be evaluated without the knowledge of the full fitness landscape. In order to
give a concrete example, we apply both the exact expression and the
approximations to a known network of over 50,000 RNA sequences. For this
neutral network, we also discuss how the fixation probability changes if
multiple sequences invade at the same time.  \bigskip

\centerline{THEORETICAL FRAMEWORK}
\bigskip

\noindent
For a population evolving under high mutational pressure, we have to
understand fixation in the sense that a mutant is fixed once it has become a
common ancestor of the whole population. The more traditional definition of
fixation, which is to regard a mutation as fixed if all sequences in the
population carry it, is not applicable: The mutational pressure constantly
creates new deleterious mutants, which may not carry a particular mutation
although their ancestors did so. If we understand fixation as the process by
which a mutant becomes a common ancestor of the whole population, then the
probability that a mutant is fixed is given by the probability that
the cascade of further mutated offspring of the invading mutant does
not come to a halt. We can calculate this probability from the theory of
multi-type branching processes.

The general setting to which our theory applies is as follows.
Consider a viral quasispecies in mutation-selection balance, with an average
fitness $\fave$. If generations are discrete and non-overlapping, and the
population size $N$ is constant, then the probability that a virion $i$
produces $k$ offspring in one generation is given by Wright-Fisher sampling,
\begin{equation}
  P(k|i) = \binom{N}{k} \xi_i^k (1-\xi_i)^{N-k}\,,
\end{equation}
with $\xi_i=w_i/(\fave N)$, where $w_i$ is the fitness of virion $i$.

Assume that a rare mutation leads to the emergence of a virion with the
potential to form a new quasispecies, and to replace the already established
one in the process. This new quasispecies (in the following also called the
invading quasispecies) may consist of sequences of type $1, 2, \dots, n$, with
replication rates $w_i$.  Let the probability that a sequence $j$ produces an
erroneous copy $i$ be given by $Q_{ij}$. As long as the total abundance of the
invading quasispecies is small compared to the established quasispecies, we
can assume that $\fave$ is not affected by the presence of the invading
quasispecies. Then, the probability that a single sequence of type $i$
generates $(k_1, \dots, k_n)$ offspring of types 1, \dots, $n$ can be
expressed as
(see Appendix)
\begin{multline}
  P(k_1, \dots, k_n|i) = \frac{N!}{(N-\sum_r k_r)!\prod_r k_r!}\\
  \times \prod_{r=1}^n
  (M_{ir}/N)^{k_r}\bigg(1-\sum_{r=1}^n
  M_{ir}/N\bigg)^{N-\sum_r k_r}\,,
\end{multline}
with $M_{ij}=w_i Q_{ji}/\fave$. The matrix elements $M_{ij}$ give the expected number of offspring of type $j$
from sequences of type $i$ in one generation.
In the following, we will assume that the population size is so large that we
can approximate  $P(k_1, \dots, k_n|i)$ by its limit for an infinitely large
population. This limit is a multivariate Poisson distribution:
\begin{equation}\label{eq:mult_poisson}
   P(k_1, \dots, k_n|i) =  \prod_{r=1}^n \left( \frac{1}{k_r!}
   M_{ir}^{k_r}\right) e^{-\sum_r M_{ir}}\,.
\end{equation}
By using the theory of branching processes, and by assuming an infinite
population size in Eq.~\eqref{eq:mult_poisson}, we restrict the applicability
of our theory to certain scenarios. We can apply our theory only to those
types of fixation events that increase the average fitness of the population.
The situation of genetic drift, whereby a neutral or deleterious mutant is
fixed because of stochastic fluctuations in a small population
\cite{Kimura70,KimuraKing79}, is not covered by our theory. This latter type
of fixation events reduces the average fitness or leaves it unaltered.

Let $x_i$ be the probability that the offspring cascade spawned by a sequence
$i$ goes extinct after a finite number of generations. From the theory of
multitype branching processes \cite{Harris63}, we know that the vector of
extinction probabilities $\vec x = (x_1, \dots, x_n)$ satisfies 
$\vec x=\vec f(\vec x)$, where $\vec f(\vec z)=\big(f_1(\vec z),
\dots, f_n(\vec z)\big)$ is the probability-generating function of the
distribution of offspring probabilities $P(k_1, \dots, k_n|i)$.
The probability-generating function is defined as 
\begin{equation}\label{eq:gen_fun}
  f_i(\vec z) = \sum_{k_1,\dots,k_n} P(k_1, \dots, k_n|i)\, z_1^{k_1}\dots z_n^{k_n}\,.
\end{equation}
After inserting Eq.~\eqref{eq:mult_poisson} into Eq.~\eqref{eq:gen_fun},
we obtain
$f_i(\vec z) = e^{\sum_{r} M_{ir}(z_r-1)}$.
With the convention $e^{\vec x}=(e^{x_1},\dots,e^{x_n})$, we can rewrite this
expression as
\begin{equation}\label{eq:gen-fun}
  \vec f(\vec z) = e^{\mat M(\vec z - 1)}\,.
\end{equation}
Since the probability of fixation $\pi_i$ of a sequence $i$ is
given by the probability that the offspring cascade spawned by $i$ does not go
extinct, we have $\pi_i=1-x_i$. The vector of fixation probabilities satisfies
therefore $1-\vec\pi=\vec f(1-\vec
\pi)$. With Eq.~\eqref{eq:gen-fun}, we find
\begin{equation}\label{eq:formal-pi-sol}
  1-\vec\pi =e^{-\mat M \vec \pi}\,.
\end{equation}
This equation has exactly one solution with $0<\pi_i<1$ for all $i$ if the
spectral radius\footnote{For the matrices $\mat M$ we are considering here,
  the spectral radius coincides with the largest positive eigenvalue of $\mat
  M$, by virtue of the Frobenius-Perron theorem.} $\rho_M$ of $\mat M$ is
larger than one \cite{Harris63}. Otherwise, $\pi_i=0$
for all $i$.

In order to compare Eq.~\eqref{eq:formal-pi-sol} to the result of Haldane
\citeyear{Haldane27},
we take
the logarithm on both sides of Eq.~\eqref{eq:formal-pi-sol} and expand to
second order:
\begin{equation}
  \log(1-\pi_i) \approx -\pi_i-\pi_i^2/2 = -\sum_kM_{ik}\pi_k\,.
\end{equation}
With $s_i=M_{ii}-1$, this simplifies to
\begin{equation}\label{eq:pi-approx}
  \pi_i = s_i+\sqrt{s_i^2+2\sum_{k\neq i} M_{ik}\pi_k}\,.
\end{equation}
If $s_i>0$ and all off-diagonal elements of $\mat M$ are zero, then
Eq.~\eqref{eq:pi-approx} reduces to Haldane's result $\pi_i=2s_i$, that is, the
fixation probability of a sequence is twice its selective advantage. If the
off-diagonal elements are non-zero, then the fixation probability is
increased, because the invading sequence gets support from its mutational
neighbors. In particular, even if some $s_i<0$, the corresponding $\pi_i$ are
positive as long as $\rho_M>1$. This means that in quasispecies fixation,
sequences that by themselves reproduce too slowly to outcompete the currently
established quasispecies can nevertheless found a new quasispecies that grows
fast enough to overtake the population.

For simplicity, we have considered only discrete, non-overlapping generations.
Generalization to continuous time is straightforward, see
e.g. \shortcite{Hermissonetal2002,Harris63}.
%%%Change
In the continuous time case, the vector of fixation probabilities $\vec\pi$ is
again determined by an equation of the form $1-\vec\pi=\vec f(1-\vec\pi)$.
However, the generating function $\vec f(\vec z)$ is in general not given by
Eq.~\eqref{eq:gen-fun}. Its functional form depends on the details of the
continuous time process that is being modeled. For example, if reproduction
occurs through binary fission, $\vec f(\vec z)$ will be quadratic in
the variables $z_1,\dots z_n$.
%%%EndChange
\bigskip

\centerline{FIXATION ON A NEUTRAL NETWORK}
\bigskip

\noindent
\textbf{Exact expressions and estimates}

\noindent
So far, we have made no assumptions about the structure and fitness
distribution of the invading quasispecies. This has led to a general equation
for the vector of fixation probabilities $\vec \pi$, but not much further
analysis is possible without a concrete model for the fitness landscape of the
invading quasispecies (we do not have to make any further assumptions about
the established quasispecies, since it enters the equations only through its
average fitness $\fave$). The concrete fitness landscape we study is that of a
neutral network
\shortcite{Huynenetal96,BornbergBauer97}
of related sequences with
identical replication rate $\sigma$. All sequences that are not part of the
neutral network are assumed to have a vanishing replication rate. Mutations
occur as random substitutions of single bases, and we allow for at most one
substitution per replication event, similar to the approach of van Nimwegen et al.
\citeyear{vanNimwegenetal99b}.
The
probability of a substitution is given by $\mu$. The restriction to at most a
single substitution is a technicality that simplifies the analysis.
Generalization to more elaborate mutation schemes is possible along the lines
of \cite{Wilke2001a}.

We denote the sequence length by $L$, and the number of different bases by
$\kappa$ ($\kappa=4$ for RNA/DNA).
For the matrix $\mat M$, we only have to take into account the sequences
belonging to the neutral network. It is useful to introduce the connection
graph $\mat G=(G_{ij})$. The elements $G_{ij}$ are 1 if and only if two
sequences $i$ and $j$ are exactly one mutation apart. In all other cases,
$G_{ij}=0$. We can express $\mat M$ in terms of $\mat G$ as
\begin{equation}\label{M_for_neutral_net}
  \mat M = (s+1)\Idmat + \beta\mat G\,,
\end{equation}
where $s=\sigma(1-\mu)/\fave-1$, $\beta=\sigma\mu/[\fave L(\kappa-1)]$,
and $\Idmat$ is the identity matrix. We restrict our analysis to \emph{primitive} connection graphs, in which case the spectral radius $\rho_G$
of $\mat G$ is given by the unique positive eigenvalue of largest modulus of
$\mat G$
\cite{Varga2000}.
(Irreducibility, which is often assumed in similar
contexts, is not sufficient, since complex eigenvalues of modulus $\rho_G$ may
exist if $\mat G$ is not primitive. Irreducible undirected connection graphs
of the kind we are considering here are primitive if they contain at least one cycle
of odd length.)

The spectral radius of $\mat M$ is given
in terms of the spectral radius of the connection graph $\rho_G$ as
\begin{equation}\label{eq:rhoMofrhoG}
  \rho_M = s+1+\beta\rho_G\,.
\end{equation}
This implies that fixation can occur as long as $s$ is not smaller than $-\beta\rho_G$.

In an experimental setting, we cannot expect to have knowledge of the complete
connection graph $\mat G$. Therefore, it is important to have approximations
for the fixation probability $\pi_i$.  We consider two alternative methods.
Both are based on replacing the matrix $\mat M$ in
Eq.~\eqref{eq:formal-pi-sol} by a suitable diagonal matrix. This replacement
leads to a decoupling of the equations for different $\pi_i$.

The quantity that is easiest to obtain experimentally is the growth rate of the
invading quasispecies relative to the established quasispecies, when initially
both are present in large and equal amounts. From the definition of $\mat M$,
we see that this relative growth rate corresponds to the spectral radius
$\rho_M$  of
$\mat M$. If we assume that every mutant present in the invading
quasispecies has an expectation of $\rho_M$ offspring per generation, then we
can replace $\mat M$ in Eq.~\eqref{eq:formal-pi-sol} with a matrix that has
entries $\rho_M$ on the diagonal, while all off-diagonal elements are
zero. Then, Eq.~\eqref{eq:formal-pi-sol} simplifies to $1-\pi_i = e^{-\rho_M
  \pi_i}$ for all $i$. Clearly, this approximation will overestimate the
$\pi_i$ for some mutants (mostly those that produce on average less than
$\rho_M$ offspring) and underestimate it for others (mostly those that produce
on average more than $\rho_M$ offspring). In the following, we will refer to
this estimate as the \emph{deterministic growth} estimate, because it is based
on the assumption that the invading quasispecies grows according to the
deterministic equations from the outset.

The alternative method of estimating $\pi_i$ is as follows.  It is
reasonable to assume that the first couple of replication cycles mostly
determine fixation or extinction for an invading sequence.  During these
initial generations, the subpopulation descending from the invading sequence
cannot explore the full neutral network if the network is large. Therefore,
the major contribution to the fixation probability comes from the connection
matrix of the local genetic neighborhood of the invading sequence, and
sequences further away on the neutral network are relatively unimportant. The
idea behind the second approximation is therefore to calculate the fixation
probability based on a small area of genotype space surrounding the invading
sequence. In the simplest case, we consider only the invading sequence and its
immediate mutational neighbors.  Assume sequence $i$ has $\nu_i$ neutral
neighbors, i.e., $\sum_j G_{ij} = \nu_i$. Then the total expected number of
offspring of sequence $i$ is $\sum_j M_{ij}=s+1+\beta\nu_i$. Under the
assumption that all offspring of $i$ have the same expected number of further
offspring, the probability of fixation satisfies the equation
$1-\pi_i=e^{-(s+1+\beta\nu_i)\pi_i}$. We call the solution to this equation
the \emph{neutrality} estimate. As in the case of the deterministic growth
estimate, it will overestimate the true fixation probability for some
sequences, and underestimate it for others.
\bigskip

\noindent
\textbf{Fixation on a RNA neutral network}

\noindent
We compared the two estimates to the exact fixation probabilities on a neutral
network of RNA sequences. The network of 51,028 sequences of length $L=18$ was
found through exhaustive enumeration by van Nimwegen et
al. \citeyear{vanNimwegenetal99b}.
The spectral radius of the network's connection
graph is $\rho_G = 15.7$.
%%%Change
In order to calculate fixation probabilities on this neutral network, we have
to make an assumption about the average fitness $\fave$ of the established
quasispecies. We assume $\fave = 1-\mu[1-\rho_G/(3L)]$, in which case the
relative growth rate of the invading quasispecies (at macroscopic
concentration) with respect to the established quasispecies follows from
Eq.~\eqref{eq:rhoMofrhoG} as
$\rho_M=\sigma$, independent of the mutation rate.
%%%EndChange

Figure~\ref{fig:exact-vs-approx} displays the exact fixation probabilities
%%%Change
(obtained numerically from Eq.~\ref{eq:formal-pi-sol}) and
%%%EndChange
the two estimates as functions of the mutation rate.  We have shown the
average fixation probability $\piave = \sum\pi_i/n$, the minimum probability
$\pi_{\min}=\min_i\{\pi_i\}$, and the maximum probability
$\pi_{\max}=\max_i\{\pi_i\}$. Since we chose $\fave$ such that $\rho_M$ is
independent of $\mu$, the deterministic growth estimate is independent of
$\mu$. We observe that the deterministic growth estimate lies consistenly
above the average $\piave$, but below the maximum $\pi_{\max}$. The neutrality
estimate underestimates the smallest fixation probabilities and overestimates
the largest ones. Its average lies slightly below $\piave$ for small mutation
rates, and above $\piave$ for large mutation rates. A more detailed plot of
the fixation probabilities at a fixed mutation rate of $\mu=0.5$ is given in
Fig.~\ref{fig:fixprob-vs-neutrality}. There, we display the fixation
probability versus the neutrality (number of neutral neighbors) of the
invading sequence. The spread in the fixation probabilities is remarkable. For
sequences with a given neutrality, the fixation probabilities vary over up to
seven orders of magnitude. This demonstrates the important influence of not
only the nearest neighbors, but also the wider genetic neighborhood on the
fate of a single sequence in quasispecies evolution. The neutrality estimate
substantially underestimates the fixation probabilities of those sequences
that have only few immediate neutral neighbors, but are otherwise located in a
region of the genotype space where the density of neutral sequences is high.
In principle, we could improve the neutrality estimate by taking into account
all neutral sequences up to some distance $d$, but in practice this method
becomes quickly as unwieldy as calculating the exact fixation probabilities.
\bigskip

\noindent
\textbf{Multiple invading sequences}

\noindent
The above considerations address only the case of a single invading sequence.
The generalization to more than one invading sequence is
straightforward. Assume that a set $\cal S$ of $N$ sequences, with ${\cal
  S}=\{i_1,\dots,i_N\}$, invades an established quasispecies. The probability
that this invasion is successful is given by
$1-\prod_{i\in {\cal S}}(1-\pi_i)$,
where $\pi_i$ are the fixation probabilities of the individual sequences.
The probability of successful invasion of $N$ sequences can be used as an
indicator for the population size at which the deterministic quasispecies
equations capture the relevant dynamics of a finite population. The
fluctuations distinguishing the stochastic process of a finite population from
the deterministic description can be neglected if the invasion probability is
close to one. In Fig.~\ref{fig:multi-sequ-invasion}, the fixation probability on
the same neutral network of RNA sequences that we have used before is
displayed against the size of the invading population. The individual data
points are averaged over 1000 independent trials, where for each trial the $N$
starting sequences were chosen at random. As before, $\fave$ is chosen such
that $\sigma$ is the average number of offspring of the invading quasispecies
in the deterministic limit.

Figure~\ref{fig:multi-sequ-invasion} shows that the population need not cover
the relevant sequence space in order to behave as predicted by the
deterministic equations. On a neutral network of over 50,000 sequences, a
population of about 1000 behaves deterministically at an advantage in growth
rate of only 1\%. It is important to note that this advantage has been
calculated under the assumption of an infinite population, and that
sufficiently small populations will grow substantially slower
\shortcite{vanNimwegenetal99b}.  Apparently, here a population that covers
only 2\% of the neutral network is not sufficiently small to experience this
reduction in growth rate.  \bigskip

\centerline{DISCUSSION}
\bigskip

\noindent
The exact expression for the probability of fixation in the quasispecies
context is easy to evaluate numerically if the fitnesses of all relevant
sequences are known. However, this data is normally not available for
experimental systems, and approximations have to be used. What is most easily
available experimentally is the relative rate of growth of the two
quasispecies at macroscopic concentrations, which is the basis of the
deterministic growth estimate. Since this estimate gives only a single number,
independently of the sequence actually seeding the invading quasispecies, it
does not reflect local variations in the density of viable sequences around
the invading sequence.
%%%Change
The neutrality estimate does not suffer from this shortcoming. However, it
requires the knowledge of the fitnesses of the immediate neighbors of the
invading sequence. Although experimentally tedious, these fitnesses can be
measured in principle. For example, \citeN{ElenaLenski97} generated 225 mutant
strains of the bacterium \emph{Escherichia coli} (each mutant differed from
the wild type by one, two, or three mutations), and measured the realtive
fitnesses of the mutant strains to the wild type. The mutant neighborhood of
an RNA virus can conceivably be measured in a similar manner.

The predictive power of both the deterministic growth estimate and the
neutrality estimate depends strongly on the distribution of neutral sequences
in sequence space. For example, both estimates become exact for the case of a
uniform neutral lattice, in which all sequences have exactly the same
neutrality. Furthermore, we expect the neutrality estimate to perform
particularly well in networks in which a sequence's neutrality is strongly
correlated to the neutralities of its immediate and more distant neutral
neighbors. The deterministic growth estimate, on the other hand, will yield
best results if the neutral network does not decompose into areas that are
substantially more densely or less densely connected than other
areas. However, to what extent these conditions are met in natural systems is
questionable. As we have seen in the present paper, the connection graph of a
comparatively simple neutral network---consisting of RNA sequences that are
only eighteen base pairs long---is already so heterogeneous that both
estimates fail to give an accurate prediction of the fixation probability for
a substantial fraction of sequences on that network. It is reasonable to
assume that the distribution of high-fitness sequences in sequence space for a
RNA virus that consists of several thousand bases is at least as heterogeneous
as the one in our toy RNA network, probably more so.
%%%End Change

In the present work, we have only considered the fate of a single invading
quasispecies. However, while an invading quasispecies is moving towards
fixation or extinction, another mutant, one that belongs to a quasispecies of
even higher mean fitness, may appear. The fixation probability of the first
invader will then be modulated by the dynamics of the second one and vice
versa, an effect commonly referred to as ``clonal interference''
\cite{GerrishLenski98}.  Clonal interference has been reported in experiments
with vesicular stomatitis virus \shortcite{Mirallesetal99,Mirallesetal2000}
and with the bacterium \emph{Escherichia coli} \shortcite{deVisseretal99}.
Currently, an accurate mathematical description of clonal interference for the
quasispecies case is not available.

The approach we have followed in this work cannot directly be
generalized to include clonal interference, because the assumption of a
constant background average fitness $\fave$ is not justified in the context of
two (or more) competing branching processes. 
%%%Change:
A second problem that we have to solve in a theory of quasispecies clonal
interference is the identification of advantageous mutants. Throughout the
present paper, we have used the definition that an advantageous mutant is one
that can grow into a quasispecies with higher average fitness than that of the
currently established quasispecies. In order to use this definition in the
context of clonal interference, we need to have a priori knowledge about how
to best subdivide the sequence space into independent quasispecies. Only with
this knowledge can we decide whether a particular new mutant is part of the
parent quasispecies, or rather the founding member of a new quasispecies.  A
possible way to study clonal interference in future work will be to consider a
particular fitness landscape---for example, a set of intertwined neutral
networks at different fitness levels---for which the a priori separation into
distinct quasispecies is possible. For such a landscape, numerical studies of
clonal interference will be straightforward, and an analytic description
should be possible as well. For landscapes that are a priori unknown, even the
numerical investigation of clonal interference will remain difficult until a
workable method for the identification of advantageous mutants has been found.
%%%End change

Recently, Jenkins et al. \citeyear{Jenkinsetal2001} and Holmes and Moya
\citeyear{HolmesMoya2002} expressed doubts regarding the relevancy of the
quasispecies model for virology (but see \citeNP{Domingo2002}). They argued
that there is no unequivocal experimental evidence for the quasispecies nature
of RNA viruses, and that the deterministic quasispecies equations are
potentially not applicable to viral evolution on theoretical grounds, due to
the immense size of the sequence space. The results of the present paper show
that the second concern is not entirely justified. A single sequence has a
positive probability to rise to fixation if and only if the average fitness of
the quasispecies that will form eventually exceeds the average fitness of the
currently established quasispecies. The individual fitness of the invading
sequence has some influence on the exact value of that probability, but does
not affect whether fixation is possible at all. Moreover, when the population
size reaches several hundred, with probability of almost one the population
will, for reasonable choices of the parameters, behave as predicted by the
deterministic equations. A similar result has been obtained by
\shortcite{vanNimwegenetal99b} for flow reactor simulations, where on the same
neutral network of RNA sequences that we have studied here, quasispecies
effects started to become important when the product of population size and
mutation rate $N\mu$ exceeded the value 10 [see Fig.~3 of
\shortcite{vanNimwegenetal99b}].

Wilke \citeyear{Wilke2001b} studied the probability of fixation for RNA
sequences in a simulated flow reactor. The measured fixation probability was
compared to an expression equivalent to the deterministic growth estimate of
the present work (since continuous time simulations were used to generate the
data, the exact expressions differ from those given here). The analytic
expression correctly predicted the parameter regions for which fixation was
possible. In particular, the mutation rate at which a slower replicator with
better mutational support could successfully invade a quasispecies consisting
of sequences with higher individual fitnesses was determined accurately.
However, the exact fixation probabilities seemed to be slightly overestimated.
(Within the statistical accuracy of the data, a definite decision on this
issue could not be made. While the data was in agreement with the model
according to a $\chi^2$ test, it was not in agreement according to a
non-parametric test based on how often the data points fell above or below the
predicted value.)

The probability of fixation of advantageous mutants is obviously of tremendous
importance for disease dynamics and vaccines. For example, live vaccinces of
attenuated poliovirus can contain small amounts of virulent poliovirus
variants \shortcite{Chumakovetal91}, the reason being that attenuated and
virulent virus variants are often separated by only one or a few mutations.
In experiments, small amounts of highly virulent virus remain typically
suppressed by the less virulent virus, but once a threshold concentration of
the highly virulent virus variant is reached, infection occurs
\shortcite{delaTorreHolland90,Chumakovetal91,Tengetal96}.  The apparent
existence of such a threshold may well be a result of insufficient resolution
of the experiments. Whether the highly virulent strain will grow is determined
by stochastic fluctuations, and as we have seen in
Fig.~\ref{fig:multi-sequ-invasion}, the probability of fixation decays quickly
with shrinking initial concentration of the virulent strain. If such a 
strain in a vaccine has a 1\% chance to cause infection, then well over a
hundred replicates of the appropriate assay are necessary to observe at least
one infection with certainty.  Probabilities of this magnitude or lower can
easily be missed at low numbers of replicates, so that the virulent strain
appears to be safely suppressed.  \bigskip

\centerline{ACKNOWLEDGMENTS}
\bigskip

\noindent
I thank C. Adami for extensive discussions and encouragement, and E. C. Holmes
for commenting on an earlier version of this manuscript.  Moreover, I thank M.
Huynen for providing the neutral network data from
\shortcite{vanNimwegenetal99b}.  Financial support by the NSF under contract
No DEB-9981397 is gratefully acknowledged.

%\bibliographystyle{chicago}
%\bibliography{paper.bib}

\begin{thebibliography}{}

\bibitem[\protect\citeauthoryear{Barton}{Barton}{1995}]{Barton95}
Barton, N.~H., 1995
\newblock Linkage and the limits to natural selection.
\newblock { Genetics\/}~{\bf 140}: 821-841.

\bibitem[\protect\citeauthoryear{Biebricher and Luce}{Biebricher and
  Luce}{1993}]{BiebricherLuce93}
Biebricher, C.~K., and R.~Luce, 1993
\newblock Sequence analysis of {RNA} species synthesized by {Q$\beta$}
  replicase without template.
\newblock { Biochemistry\/}~{\bf 32}: 4848-4854.

\bibitem[\protect\citeauthoryear{Bornberg-Bauer}{Bornberg-Bauer}{1997}]{Bornbe%
rgBauer97}
Bornberg-Bauer, E., 1997
\newblock How are model protein structures distributed in sequence space?
\newblock { Biophys. Journal\/}~{\bf 73}: 2393-2403.

\bibitem[\protect\citeauthoryear{Burch and Chao}{Burch and
  Chao}{2000}]{BurchChao2000}
Burch, C.~L., and L.~Chao, 2000
\newblock Evolvability of an {RNA} virus is determined by its mutational
  neighbourhood.
\newblock { Nature\/}~{\bf 406}: 625-628.

\bibitem[\protect\citeauthoryear{B\"urger and Ewens}{B\"urger and
  Ewens}{1995}]{BurgerEwens95}
B\"urger, R., and W.~J. Ewens, 1995
\newblock Fixation probabilities of additive alleles in diploid populations.
\newblock { J. Math. Biol.\/}~{\bf 33}: 557-575.

\bibitem[\protect\citeauthoryear{Chumakov, Powers, Noonan, Roninson, and
  Levenbook}{Chumakov et~al.}{1991}]{Chumakovetal91}
Chumakov, K.~M., L.~B. Powers, K.~E. Noonan, I.~B. Roninson, and I.~S.
  Levenbook, 1991
\newblock Correlation between amount of virus with altered nucleotide sequence
  and the monkey test for acceptability of oral poliovirus vaccine.
\newblock { Proc. Natl. Acad. Sci. USA\/}~{\bf 88}: 199-203.

\bibitem[\protect\citeauthoryear{{de la Torre} and Holland}{{de la Torre} and
  Holland}{1990}]{delaTorreHolland90}
{de la Torre}: J.~C., and J.~J. Holland, 1990
\newblock {RNA} virus quasispecies populations can suppress vastly superior
  mutant progeny.
\newblock { J. Virol.\/}~{\bf 64}: 6278-6281.

\bibitem[\protect\citeauthoryear{de~Visser, Zeyl, Gerrish, Blanchard, and
  Lenski}{de~Visser et~al.}{1999}]{deVisseretal99}
de~Visser, J. A. G.~M., C.~W. Zeyl, P.~J. Gerrish, J.~L. Blanchard, and R.~E.
  Lenski, 1999
\newblock Diminishing returns from mutation supply rate in asexual populations.
\newblock { Science\/}~{\bf 283}: 404-406.

\bibitem[\protect\citeauthoryear{Demetrius, Schuster, and Sigmund}{Demetrius
  et~al.}{1985}]{Demetriusetal85}
Demetrius, L., P.~Schuster, and K.~Sigmund, 1985
\newblock Polynucleotide evolution and branching processes.
\newblock { Bull. Math. Biol.\/}~{\bf 47}: 239-262.

\bibitem[\protect\citeauthoryear{Domingo}{Domingo}{2002}]{Domingo2002}
Domingo, E., 2002
\newblock Quasispecies theory in virology.
\newblock { J. Virol.\/}~{\bf 76}: 463-465.

\bibitem[\protect\citeauthoryear{Domingo, Biebricher, Eigen, and
  Holland}{Domingo et~al.}{2001}]{Domingoetal2001}
Domingo, E., C.~K. Biebricher, M.~Eigen, and J.~J. Holland, 2001
\newblock {\em Quasispecies and {RNA} Virus Evolution: Principles and
  Consequences}.
\newblock Georgetown, TX: Landes Bioscience.

\bibitem[\protect\citeauthoryear{Domingo, Flavell, and Weissmann}{Domingo
  et~al.}{1976}]{Domingoetal76}
Domingo, E., R.~A. Flavell, and C.~Weissmann, 1976
\newblock \emph{In vitro} site directed mutagenesis: generation and properties
  of an infectious extracistronic mutant of bacteriophage {Q}$\beta$.
\newblock { Gene\/}~{\bf 1}: 3-25.

\bibitem[\protect\citeauthoryear{Domingo and Holland}{Domingo and
  Holland}{1997}]{DomingoHolland97}
Domingo, E., and J.~J. Holland, 1997
\newblock {RNA} virus mutations and f\/itness for survival.
\newblock { Annu. Rev. Microbiol.\/}~{\bf 51}: 151-178.

\bibitem[\protect\citeauthoryear{Domingo, Sabo, Taniguchi, and
  Weissmann}{Domingo et~al.}{1978}]{Domingoetal78}
Domingo, E., D.~Sabo, T.~Taniguchi, and C.~Weissmann, 1978
\newblock Nucleotide sequence heterogeneity of an {RNA} phage population.
\newblock { Cell\/}~{\bf 13}: 735-744.

\bibitem[\protect\citeauthoryear{Drake}{Drake}{1993}]{Drake93}
Drake, J.~W., 1993
\newblock Rates of spontaneous mutation among {RNA} viruses.
\newblock { Proc. Natl. Acad. Sci. USA\/}~{\bf 90}: 4171-4175.

\bibitem[\protect\citeauthoryear{Drake and Holland}{Drake and
  Holland}{1999}]{DrakeHolland99}
Drake, J.~W., and J.~J. Holland, 1999
\newblock Mutation rates among {RNA} viruses.
\newblock { Proc. Natl. Acad. Sci. USA\/}~{\bf 96}: 13910-13913.

\bibitem[\protect\citeauthoryear{Eigen and Schuster}{Eigen and
  Schuster}{1979}]{EigenSchuster79}
Eigen, M., and P.~Schuster, 1979
\newblock {\em The Hypercycle - A Principle of Natural Self-Organization}.
\newblock Berlin: Springer-Verlag.

\bibitem[\protect\citeauthoryear{Elena and Lenski}{Elena and
  Lenski}{1997}]{ElenaLenski97}
Elena S.~F., and R.~E. Lenski, 1997
\newblock Test of synergistic interactions among deleterious mutations in bacteria.
\newblock { Nature\/}~{\bf 390}: 395-398.

\bibitem[\protect\citeauthoryear{Ewens}{Ewens}{1967}]{Ewens67}
Ewens, W.~J., 1967
\newblock The probability of f\/ixation of a mutant: the two-locus case.
\newblock { Evolution\/}~{\bf 21}: 532-540.

\bibitem[\protect\citeauthoryear{Fisher}{Fisher}{1922}]{Fisher22}
Fisher, R.~A., 1922
\newblock On the dominance ratio.
\newblock { Proc. Roy. Soc. Edin.\/}~{\bf 42}: 321-341.

\bibitem[\protect\citeauthoryear{Fisher}{Fisher}{1930}]{Fisher30}
Fisher, R.~A., 1930
\newblock The distribution of gene ratios for rare mutations.
\newblock { Proc. Roy. Soc. Edin.\/}~{\bf 50}: 204-219.

\bibitem[\protect\citeauthoryear{Gerrish and Lenski}{Gerrish and
  Lenski}{1998}]{GerrishLenski98}
Gerrish, P.~J., and R.~E. Lenski, 1998
\newblock The fate of competing benef\/icial mutations in an asexual population.
\newblock { Genetica\/}~{\bf 102/103}: 127-144.

\bibitem[\protect\citeauthoryear{Haldane}{Haldane}{1927}]{Haldane27}
Haldane, J. B.~S., 1927
\newblock A mathematical theory of natural and artif\/icial selection. {P}art
  {V}: {S}election and mutation.
\newblock { Proc. Camp. Phil. Soc.\/}~{\bf 23}: 838-844.

\bibitem[\protect\citeauthoryear{Harris}{Harris}{1963}]{Harris63}
Harris, T.~E., 1963
\newblock {\em The Theory of Branching Processes}.
\newblock Berlin: Springer.

\bibitem[\protect\citeauthoryear{Hermisson, Redner, Wagner, and
  Baake}{Hermisson et~al.}{2002}]{Hermissonetal2002}
Hermisson, J., O.~Redner, H.~Wagner, and E.~Baake, 2002
\newblock Mutation-selection balance: Ancestry, load, and maximum principle.
\newblock { Theor. Pop. Biol.\/}.
\newblock in press.

\bibitem[\protect\citeauthoryear{Hofbauer and Sigmund}{Hofbauer and
  Sigmund}{1988}]{HofbauerSigmund88}
Hofbauer, J., and K.~Sigmund, 1988
\newblock {\em The Theory of Evolution and Dynamical Systems}.
\newblock Cambridge: Cambridge University Press.

\bibitem[\protect\citeauthoryear{Holland, Spindler, Horodyski, Grabau, Nichol,
  and VandePol.}{Holland et~al.}{1982}]{Hollandetal82}
Holland, J., K.~Spindler, F.~Horodyski, E.~Grabau, S.~Nichol, and S.~VandePol., 1982
\newblock Rapid evolution of {RNA} genomes.
\newblock { Science\/}~{\bf 215}: 1577-1585.

\bibitem[\protect\citeauthoryear{Holmes and Moya}{Holmes and
  Moya}{2002}]{HolmesMoya2002}
Holmes, E.~C., and A.~Moya, 2002
\newblock Is the quasispecies concept relevant to {RNA} viruses?
\newblock { J. Virol.\/}~{\bf 76}: 460-462.

\bibitem[\protect\citeauthoryear{Huynen, Stadler, and Fontana}{Huynen
  et~al.}{1996}]{Huynenetal96}
Huynen, M.~A., P.~F. Stadler, and W.~Fontana, 1996
\newblock Smoothness within ruggedness: {T}he role of neutrality in adaptation.
\newblock { Proc. Natl. Acad. Sci. USA\/}~{\bf 93}: 397-401.

\bibitem[\protect\citeauthoryear{Jenkins, Worobey, Woelk, and Holmes}{Jenkins
  et~al.}{2001}]{Jenkinsetal2001}
Jenkins, G.~M., M.~Worobey, C.~H. Woelk, and E.~C. Holmes, 2001
\newblock Evidence for the non-quasispecies evolution of {RNA} viruses.
\newblock { Mol. Biol. Evol.\/}~{\bf 18}: 987-994.

\bibitem[\protect\citeauthoryear{Kimura}{Kimura}{1957}]{Kimura57}
Kimura, M., 1957
\newblock Some problems of stochastic processes in genetics.
\newblock { Ann. Math. Statist.\/}~{\bf 28}: 882-901.

\bibitem[\protect\citeauthoryear{Kimura}{Kimura}{1964}]{Kimura64}
Kimura, M., 1964
\newblock Dif\/fusion models in population genetics.
\newblock { J. Appl. Prob.\/}~{\bf 1}: 177-232.

\bibitem[\protect\citeauthoryear{Kimura}{Kimura}{1970}]{Kimura70}
Kimura, M., 1970
\newblock The length of time required for a selectively neutral mutant to
reach fixation through random frequency drift in a finite population.
\newblock { Genetical Research\/}~{\bf 15}: 131-133.

\bibitem[\protect\citeauthoryear{Kimura and King }{Kimura and King}{1979}]{KimuraKing79}
Kimura, M., and J.~L. King, 1979
\newblock Fixation of a deleterious allele at one of two ``duplicate'' loci by
mutation pressure and random drift.
\newblock { Proc. Natl. Acad. Sci. USA\/}~{\bf 76}: 2858-2861.

\bibitem[\protect\citeauthoryear{Krakauer and Plotkin}{Krakauer and
  Plotkin}{2002}]{KrakauerPlotkin2002}
Krakauer, D.~C., and J.~B. Plotkin, 2002
\newblock Redundancy, antiredundancy, and the robustness of genomes.
\newblock { Proc. Natl. Acad. Sci. USA\/}~{\bf 99}: 1405-1409.

\bibitem[\protect\citeauthoryear{Miralles, Gerrish, Moya, and Elena}{Miralles
  et~al.}{1999}]{Mirallesetal99}
Miralles, R., P.~J. Gerrish, A.~Moya, and S.~F. Elena, 1999
\newblock Clonal interference and the evolution of rna viruses.
\newblock { Science\/}~{\bf 285}: 1745-1747.

\bibitem[\protect\citeauthoryear{Miralles, Moya, and Elena}{Miralles
  et~al.}{2000}]{Mirallesetal2000}
Miralles, R., A.~Moya, and S.~F. Elena, 2000
\newblock Diminishing returns of population size in the rate of {RNA} virus
  adaptation.
\newblock { J. Virol.\/}~{\bf 74}: 3566-3571.

\bibitem[\protect\citeauthoryear{Nowak}{Nowak}{1992}]{Nowak92}
Nowak, M.~A., 1992
\newblock What is a quasispecies?
\newblock { TREE\/}~{\bf 7}: 118-121.

\bibitem[\protect\citeauthoryear{Otto and Barton}{Otto and
  Barton}{1997}]{OttoBarton97}
Otto, S.~P., and N.~H. Barton, 1997
\newblock The evolution of recombination: removing the limits to natural
  selection.
\newblock { Genetics\/}~{\bf 147}: 879-906.

\bibitem[\protect\citeauthoryear{Pollak}{Pollak}{2000}]{Pollak2000}
Pollak, E., 2000
\newblock Fixation probabilities when the population size undergoes cyclic
  f\/luctuations.
\newblock { Theor. Pop. Biol.\/}~{\bf 57}: 51-58.

\bibitem[\protect\citeauthoryear{Schuster and Swetina}{Schuster and Swetina}{1988}]{SchusterSwetina88}
Schuster, P., and J. Swetina, 1988
\newblock Stationary mutant distributions and evolutionary optimization
\newblock { Bull. Math. Biol.\/}~{\bf 50}: 635-660.

\bibitem[\protect\citeauthoryear{Steinhauer, {de la Torre}, Meier, and
  Holland}{Steinhauer et~al.}{1989}]{Steinhaueretal89}
Steinhauer, D.~A., J.~C. {de la Torre}, E.~Meier, and J.~J. Holland, 1989
\newblock Extreme heterogeneity in populations of vesicular stomatitis virus.
\newblock { J. Virol.\/}~{\bf 63}: 2072-2080.

\bibitem[\protect\citeauthoryear{Teng, Oldstone, and {de la Torre}}{Teng
  et~al.}{1996}]{Tengetal96}
Teng, M.~N., M.~B.~A. Oldstone, and J.~C. {de la Torre}, 1996
\newblock Suppression of lymphocytic choriomeningitis virus - induced growth
  hormone def\/iciency syndrome by disease-negative virus variants.
\newblock { Virology\/}~{\bf 223}: 113-119.

\bibitem[\protect\citeauthoryear{van Nimwegen, Crutchfield, and Huynen}{van
  Nimwegen et~al.}{1999}]{vanNimwegenetal99b}
van Nimwegen, E., J.~P. Crutchf\/ield, and M.~Huynen, 1999
\newblock Neutral evolution of mutational robustness.
\newblock { Proc. Natl. Acad. Sci. USA\/}~{\bf 96}: 9716-9720.

\bibitem[\protect\citeauthoryear{Varga}{Varga}{2000}]{Varga2000}
Varga, R.~S., 2000
\newblock {\em Matrix Iterative Analysis\/} (2nd ed.).
\newblock New York: Springer Verlag.

\bibitem[\protect\citeauthoryear{Wilke}{Wilke}{2001a}]{Wilke2001a}
Wilke, C.~O., 2001a
\newblock Adaptive evolution on neutral networks.
\newblock { Bull. Math. Biol.\/}~{\bf 63}: 715-730.

\bibitem[\protect\citeauthoryear{Wilke}{Wilke}{2001b}]{Wilke2001b}
Wilke, C.~O., 2001b
\newblock Selection for f\/itness versus selection for robustness in {RNA}
  secondary structure folding.
\newblock { Evolution\/}~{\bf 55}: 2412-2420.

\bibitem[\protect\citeauthoryear{Wilke, Wang, Ofria, Lenski, and Adami}{Wilke
  et~al.}{2001}]{Wilkeetal2001b}
Wilke, C.~O., J.~L. Wang, C.~Ofria, R.~E. Lenski, and C.~Adami, 2001
\newblock Evolution of digital organisms at high mutation rate leads to
  survival of the f\/lattest.
\newblock { Nature\/}~{\bf 412}: 331-333.

\end{thebibliography}

%%%Change

\centerline{APPENDIX}
\bigskip

We consider a model with discrete, non-overlapping generations, and a constant
population size $N$. Under the assumption that the reproductive success of a
sequence $i$ is proportional to its fitness $w_i$, the probability that a
randomly chosen sequence in the next generation is offspring of sequence $i$
is given by $\xi=w_i/(\fave N)$, where $\fave$ is the average fitness in the
population. Since there are $N$ sequences in the population, the probability
that $k$ of them are offspring of sequence $i$ is binomial,
$P(k|i)=\binom{N}{k}\xi^k(1-\xi)^{N-k}$. Now consider a sequence of type $r$
in the offspring generation. For the probability that the parent of sequence
$r$ is a particular sequence $i$ of the previous generation, we find
$\xi_r=Q_{ri}\xi=w_iQ_{ri}/(\fave N)$, because only a fraction $Q_{ri}$ of the
total offspring of $i$ will be of type $r$. Following the previous argument,
we find for the probability that sequence $i$ leaves $k_r$ offspring of
type~$r$: $P(k_r|i)=\binom{N}{k_r}\xi_r^k(1-\xi_r)^{N-k_r}$.

We can extend the above argument to sequences of two types $r$ and $s$. The
probability that sequence $i$ leaves $k_r$ offspring sequences of type $r$ and
$k_s$ offspring sequences of type $s$ is the probability that $k_r$ offspring
are of type $r$, $\xi_r^{k_r}$, times the probability that $k_s$ offspring are
of type $s$, $\xi_s^{k_s}$, times the probability that the remaining offspring
are either of different types, or have different parent sequences,
$(1-\xi_r-\xi_s)^{N-k_r-k_s}$, times the number of possible ways in which
$k_r$ and $k_s$ sequences can be chosen out of the total of $N$ sequences in
the population. This latter number is a multinomial coefficient,
$N!/[k_r!k_s!(N-k_r-k_s)!]$. Putting everything together, we find
\begin{equation}\tag{A1}
   P(k_r, k_s|i) =  \frac{N!}{k_r!k_s!(N-k_r-k_s)!}
   \xi_r^{k_r}\xi_{s}^{k_s}(1-\xi_r-\xi_{s})^{N-k_r-k_s}\,.
\end{equation}
By repeating this argument for $n$ different sequence types, and with the
definition 
$M_{ij}:=N\xi_j=w_iQ_{ji}/\fave$, we arrive at Eq.~\eqref{eq:mult_poisson}.

%%%End Change

\clearpage
\newpage

\begin{figure}
\caption{\label{fig:exact-vs-approx}Fixation probability
  versus mutation rate in a neutral network of 51,028 RNA sequences taken
  from (van Nimwegen et al. 1999).
%  \cite{vanNimwegenetal99b}
  Solid lines correspond to the solution of the full equations, dashed lines
  correspond to the neutrality estimate, and the dotted line indicates the
  deterministic growth estimate.  ($\sigma=1.05$, $L=18$, $\rho_G=15.7$,
  $\fave = 1-\mu[1-\rho_G/(3L)]$.) }
\end{figure}

\begin{figure}
\caption{\label{fig:fixprob-vs-neutrality}Fixation probability
  versus neutrality $\nu$ of the invading sequence in a neutral network of
  51,028 RNA sequences taken from (van Nimwegen et al. 1999).
%  \cite{vanNimwegenetal99b}
  The dots stem from the exact numerical solution, the
  dashed line corresponds to the neutrality estimate, and the dotted line
  indicates the deterministic growth estimate. The inset shows the
  distribution of neutralities in the network.  ($\mu=0.5, \sigma=1.05$, $L=18$, $\rho_G=15.7$, $\fave =
  1-\mu[1-\rho_G/(3L)]$.)}
\end{figure}

\begin{figure}
\caption{\label{fig:multi-sequ-invasion}Fixation probability $\pi$
  versus size of the invading population $N$ in neutral network of 51,028 RNA
  sequences.
  The fixation probability is averaged over 1000 independent sets of invading
  sequences, chosen at random. The error bars indicate the standard deviation.
  Lines are meant as a guide to the eye. ($\mu=.2$, $L=18$, $\rho_G=15.7$, $\fave = 1-\mu[1-\rho_G/(3L)]$.)}
\end{figure}

\clearpage
\newpage
\centerline{
\includegraphics[width=.98\columnwidth]{exact-vs-approx}
}
\noindent Figure 1

\newpage
\centerline{
\includegraphics[width=.98\columnwidth]{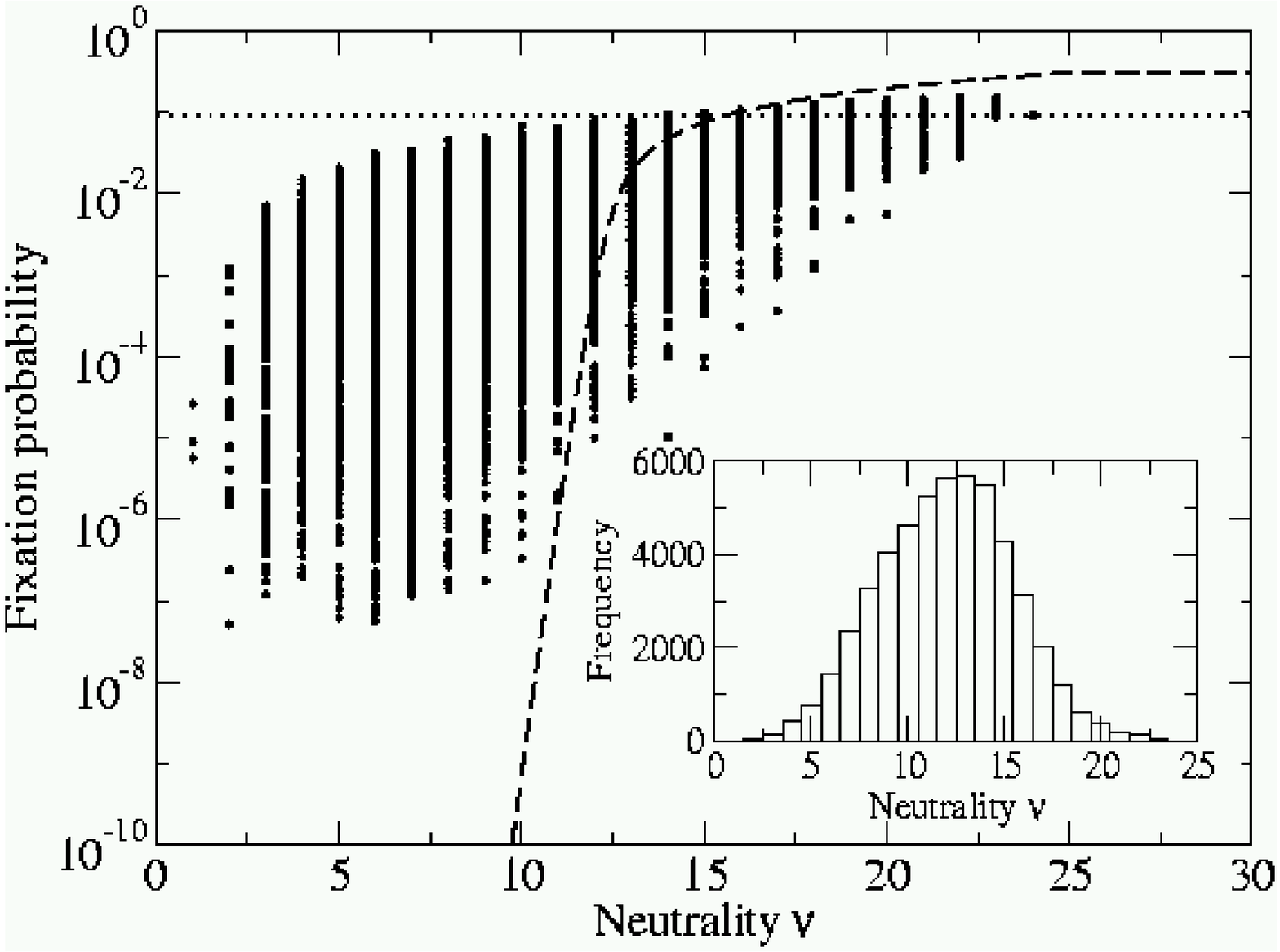}
}
\noindent Figure 2

\newpage
\centerline{
\includegraphics[width=.98\columnwidth]{multi-sequ-invasion}
}
\noindent Figure 3

\end{document}